\documentclass[twocolumn,prd,amssymb,amsmath,preprintnumbers,floatfix,aps]{revtex4-1}
\usepackage{dcolumn,graphicx,bm,psfrag}

\newcommand{\be}{\begin{equation}}
\newcommand{\ee}{\end{equation}}
\newcommand{\bear}{\begin{eqnarray}}
\newcommand{\eear}{\end{eqnarray}}
\newcommand{\ba}{\begin{array}}
\newcommand{\ea}{\end{array}}

\begin{document}
\title{\boldmath \bf \Large Hidden GeV-scale interactions of quarks} 
\author{\bf Bogdan A. Dobrescu and Claudia Frugiuele } 


\affiliation{Theoretical Physics Department, Fermilab, Batavia, Illinois, USA}

\date{\normalsize  April 24, 2014}

\begin{abstract}
We explore quark interactions mediated by new gauge bosons of masses in the 0.3 -- 50 GeV range. 
A tight upper limit on the gauge coupling of light $Z'$ bosons  
is imposed by  the anomaly cancellation conditions in conjunction with collider bounds on new charged fermions.
Limits from quarkonium decays are  model dependent, while electroweak constraints are mild.
We derive the limits for a $Z'$ boson coupled to baryon number, and then construct 
a  $Z'$ model with relaxed constraints, allowing quark couplings as large as 0.2 for a mass of a few GeV.
\end{abstract}

\vspace{.4cm}

\maketitle

{\it Introduction.}---Quarks experience all five known interactions: strong, weak, electromagnetic, Higgs and gravitational.
It behooves us to ask whether additional interactions of quarks exist, and what are the current limits on their strength. 
Experimental searches for new particles interacting with quarks have been performed at hadron colliders over the last few decades,
setting upper limits on their couplings for masses in the 50 GeV$-$3 TeV range \cite{Beringer:1900zz,Dobrescu:2013cmh}.
Smaller masses have been less intensely investigated, due to large backgrounds at hadron colliders.

Since the quarks are fermions, the mediators of quark interactions must be bosons. Restricting attention to renormalizable quantum 
field theories, elementary mediators can have spin 0 or 1. 
Due to the chiral nature of the known quarks, the light quarks may have only suppressed interactions with spin-0 mediators.
It is less clear whether large couplings may be allowed for spin-1 mediators. In this letter we study existing limits 
on the coupling of new spin-1 particles that interact with quarks, and have masses in the  $0.3-50$ GeV range.
Lighter mediators are possible, but their couplings are strongly constrained \cite{NelsonTetradis}, and precise limits are harder to derive 
for masses near the QCD scale.

Spin-1 fields are well behaved at high energies only if they are gauge bosons. While composite spin-1  particles
bound by some new dynamics may exist,  their coupling to quarks would suggest a
compositeness scale above the weak scale ($v \approx 246$ GeV); given that
spin-1 states are typically near the compositeness scale, we focus on elementary gauge bosons.
Only electrically-neutral gauge bosons with highly suppressed couplings to leptons (``leptophobic") are allowed at masses below 50 GeV.
These can be color singlets ({\it i.e.}, $Z'$ bosons) or octets. The latter are severely constrained by the running of the QCD coupling below 
$M_Z$ \cite{Krnjaic:2011ub}. Thus, leptophobic $Z'$ bosons associated with a $U(1)_z$  gauge extension 
of the Standard Model (SM) are the best candidates for mediating new relatively-strong quark interactions at low energies. 

\bigskip

{\it $Z'$ couplings to quarks.}---The renormalizable interactions of a $Z'$ boson of this type are given by 
\be 
\frac{g_z}{2}   Z'_\mu  \left(  z_{Q_j}  \, \overline Q_L^j  \gamma^\mu Q_L^j + z_{u_j} \,  \overline u_R^j  \gamma^\mu u_R^j + z_{d_j} \,  \overline d_R^j  \gamma^\mu d_R^j \right) ,
\label{couplings}
\ee
where $j$ labels the generations, $Q_L^j$ are the left-handed quark doublets, 
$u_R^j$, $d_R^j$ are the right-handed quark gauge eigenstates,
$z_{Q_j}$, $z_{u_j} $, $z_{d_j}$  are their $U(1)_z$ charges, and $g_z$ is the gauge coupling. 
Higher-dimensional $Z'$ interactions may exist \cite{Fox:2011qd}, but their effects for a $Z'$ mass $M_{Z'} \ll v$ are suppressed.

A simple charge assignment that allows quark masses and 
evades constraints from flavor-changing neutral currents (FCNC)
is $z_{Q_j}\! = \! z_{u_j} \! = \! z_{d_j} \! = \! 1/3$, {\it i.e.},  charges given by the baryon number. 
Experimental limits on $g_z$ are  loose in this case \cite{Carone:1994aa,Aranda:1998fr}, 
except for $M_{Z'}$ near the $\Upsilon$ or $J/\psi$ resonances.
We will show, however, that additional limits on light leptophobic $Z'$ bosons arise from the interplay of collider limits on new fermions 
and theoretical constraints.

\medskip

{\it Anomaly cancellation.}---Self-consistency of the theory at high energies \cite{Bardeen:1969md} requires gauge anomaly cancellation. 
Given that $U(1)_z$  must be embedded at some high scale in a non-Abelian gauge group,
the charges should be commensurate. Thus, for certain $g_z$ normalizations, the $U(1)_z$  charges are integers, 
so that finding solutions to the $[U(1)_z]^3$ anomaly cancellation is non-trivial.

Even without fermions beyond the SM, leptophobic $U(1)_z$ groups can be anomaly free, 
{\it e.g.}, when first (second) generation quarks have $U(1)_z$  charge $z_1$ ($z_2$),
and third-generation quarks have charge $-(z_1+z_2)$.  For $M_Z' \ll v$, though, 
$Z'$-induced FCNC are large unless the $Q_L^j$ charges are equal ($u_R^j$ or $d_R^j$ may have $j$-dependent charges because 
their gauge and mass eigenstates may be identical):
$K^0-\bar{K}^0$ mixing requires 
$g_z |z_{Q_3} - z_{Q_1}|  <  10^{-5}  M_{Z'}/(1 \, {\rm GeV})$, and $B^0-\bar{B}^0$ mixing imposes a slightly weaker constraint on  $g_z |z_{Q_3} - z_{Q_1}|$.
Thus, large values of $g_z$ require
$z_{Q_1} = z_{Q_2} = z_{Q_3} $. 
Then, in the absence of new fermions, the $[SU(2)_W]^2 U(1)_z$ anomaly implies $z_{Q_j} = 0$.
The remaining anomaly cancellations \cite{Carena:2004xs} imply  $z_{u_3} = -  z_{u_1} \!\!-\! z_{u_2}$ and $z_{d_3} = -  z_{d_1}\!\! -\! z_{d_2}$, as well as
\bear
2 ( z_{u_1}^2 + z_{u_2}^2 + z_{u_1} z_{u_2}) &=& z_{d_1}^2 + z_{d_2}^2 + z_{d_1} z_{d_2}   ~~,
\nonumber \\ [2mm]
- z_{u_1} z_{u_2} (z_{u_1} +z_{u_2}) &=& z_{d_1} z_{d_2}  ( z_{d_1} + z_{d_2} )    ~~.
\eear
A necessary condition for these equations to have integer solutions is that there exist an integer $k$ such that 
\be
z_{u_1}^3 - 6 z_{u_1}^2 z_{d_1} + 4 z_{d_1}^3  = k^2 (z_{u_1} + 2 z_{d_1})  ~~.
\ee
We have checked numerically that this condition is not satisfied for   $|z_{u_1}|,|z_{d_1}| \leq 1000$.
For practical purposes, thus,  anomaly-free solutions with generation-independent $z_{Q_j}$ require fermions beyond the SM.

A fourth generation of chiral fermions is ruled out by direct searches for new quarks at the LHC \cite{Chatrchyan:2013uxa}.
Anomaly-free sets of color-singlet chiral fermions 
\cite{deGouvea:2012hc} are severely constrained by  $h^0\to \gamma\gamma$ and electroweak measurements.

We are then lead to consider fermions that are vector-like with respect to the SM gauge group, and chiral under $U(1)_z$.
Among the new fermions required to cancel the various anomalies, there are electrically charged 
ones. If these are long-lived, then LHC searches for slowly ionizing charged tracks set a mass limit  $m_f > 450$ GeV (we have compared the 
experimental limit \cite{LongLived} with the pair-production cross section for a charge-1 lepton computed with MadGraph \cite{Alwall:2011uj}).
Decays of charged vector-like fermions into neutral ones may relax the limits. 
The 1-loop mass splitting between the components $N^0$ and $E^{\pm}$ of a weak-doublet vector-like lepton is $\sim\! 0.3$ GeV \cite{Thomas:1998wy}.
For a stable $N^0$, the process $e^+e^- \!\!\to E^+ E^-$ leads to a final state with two soft pions and missing energy.   
The mass limit, using initial state radiation at LEP, is $\sim90$ GeV \cite{Heister:2002mn}; for the future LHC reach, see \cite{Halverson:2014nwa}.

Lower limits on vector-like fermion masses translate into an upper limit on $g_z$. Note that 
$M_{Z'} = g_z z_\varphi \langle \varphi \rangle /2 $, where $\varphi$ is the scalar whose VEV breaks  $U(1)_z$. 
A new fermion, $f$, that is chiral with respect to $U(1)_z$ acquires a mass $m_f = \lambda \langle \varphi \rangle$
via a Yukawa term $\lambda \varphi \,\bar f_L f_R$. Given that the Yukawa coupling blows up in the UV, there is a perturbativity limit 
$\lambda \lesssim 4 \pi /3$; a lower limit on $m_f$ then implies a bound on $g_z$:
\be
g_z =  \frac{ 2 \lambda M_{Z'}}{z_\varphi m_f}  \lesssim
 \frac{8.4 \times 10^{-2} }{z_\varphi}  \left(\frac{M_{Z'}}{1 \; {\rm GeV}}\right)\! \left(\frac{100 \; {\rm GeV}}{m_f} \right) .
\ee

For $U(1)_z$ charges given by the baryon number (we refer to this assignment as the $Z'_B$ model), 
the $[SU(3)_c]^2 U(1)_z$ anomaly cancels, so that all 
vector-like fermions may be color singlets \cite{Carena:2004xs,Duerr:2013dza} (solutions with color triplets also exist  \cite{Dobrescu:2013cmh}). 
To avoid a large $Z$-$Z'$ mixing, the $U(1)_Y$ and $U(1)_z$ charges must satisfy Tr$(z \,Y)=0$. 
The minimal set of vector-like fermions consists of a weak doublet ($Y$=$-1/2$, $z_L$=$-1$, $z_R$=$+2$),
a weak singlet ($Y$=$-1$, $z_L$=$+2$, $z_R$=$-1$), and a SM singlet ($z_L$=$+2$, $z_R$=$-1$; or $z_L$=$+1$, $z_R$=$- 2$); 
their masses require $z_\varphi=3$.
If the charged fermions are slightly heavier than the neutral ones, then the collider signal is again soft pions and missing energy
so that $m_f > 90$ GeV, leading to the $g_z$ limit given by the middle straight line in Fig.~1.
Without tuning, though, the mass splittings are large, the collider limits on $m_f$ are higher, and the upper limit on $g_z$ decreases 
(see, {\it e.g.}, the $m_f = 450$ GeV line in Fig.~1).

A loophole is that several $\varphi$ scalars may break $U(1)_z$. If $n$ scalars
have equal VEVs and equal charges, the  $g_z$ limit is relaxed by a factor of $\sqrt{n}$. Plausible theories, though, 
do not have $n$ larger than a few. 
Another loophole is that there can be $N_f$ copies of the minimal 
vector-like fermion set with $U(1)_z$ charges smaller by a factor of $N_f$, implying $z_\varphi = 3/N_f$ and a $g_z$ limit increased by $N_f$
(see the $N_f=3$, $m_f = 90$ GeV line  in Fig. 1).

\begin{figure}[t!] 
\begin{center} \hspace*{-0.25cm}\includegraphics[width=0.475\textwidth, angle=0]{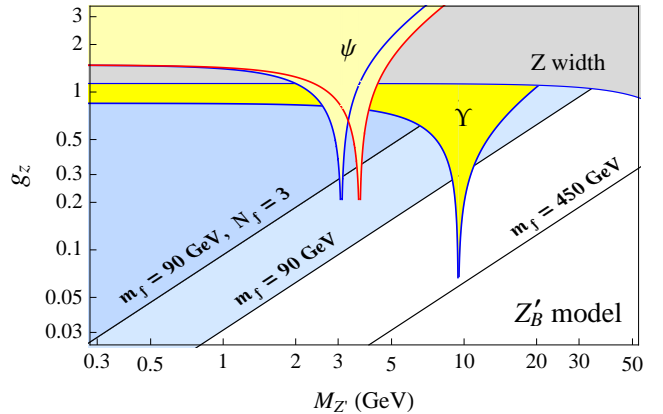}
\caption{Limits in the gauge coupling versus mass
  plane for $Z^\prime_B$. Values of $g_z$ above the straight lines are excluded by the 
  anomaly cancellation conditions in conjunction with collider searches for new 
  fermions, for $N_f = 1$ or 3, $m_f > 90$ GeV, and for  $N_f =1$, $m_f > 450$ GeV.
  The 
  top regions are excluded by quarkonium and hadronic $Z$ decays.  \\ [-8mm] } 
\label{fig:ZBlimits}
\end{center}
\end{figure}

\medskip

{\it Promptly decaying vector-like fermions.}---Let us explore whether the mass limits 
can be relaxed when the vector-like fermions decay into SM particles.
A vector-like fermion may decay through mixing with a SM one if they couple to $\varphi$. The new mass-eigenstate fermion 
has 4 decay channels, into a SM fermion and one of the heavy bosons, $W$, $Z$, $h^0$ or $Z'$. 
The branching fraction involving $Z'$ is typically small, of order
 $(g_z z_f/g)^2/4 \ll 1$, where $z_f$ is the $U(1)_z$ charge of the mixed component of the new fermion, and $g\approx 0.65$ is the weak gauge coupling.
The LHC limits on vector-like quarks that decay into a SM quark and a $W$, $Z$, or $h^0$ boson are stringent,
above 700 GeV \cite{Chatrchyan:2013uxa}.
Vector-like leptons are less constrained. The LHC limits on processes involving weak bosons and missing transverse energy or charged leptons
have been recast as mass limits on vector-like leptons \cite{Falkowski:2013jya}. 
 Weak-doublet leptons must be heavier than about 280 GeV if they decay to $\tau Z$ and $\tau W$
(the limit is 460 GeV if the $\tau$ is replaced by an $e$ or $\mu$).  

If the SM quark doublets are $U(1)_z$-neutral, the  vector-like fermions can be weak singlets, so that 
production at the LHC cannot proceed through a $W$, and the $g_z$ limit  is relaxed. 
Weak-singlet charged leptons decaying into $ e Z$ or $\mu Z$ must be heavier than $\sim100$ GeV, 
while no LHC limit can be derived for the $\tau Z$ decay \cite{Falkowski:2013jya}. 
The LEP mass limit on leptons decaying into $\nu W$ is 101 GeV \cite{Achard:2001qw}. 

Vector-like fermions lighter than $W$ would decay predominantly into a SM fermion and a $Z'$, which in turn decays into two jets.
Pair production of a vector-like quark decaying into a light quark and a $Z'$ leads to a 6-jet final state.
Masses between 77 GeV and $\sim 100$ GeV are ruled out by a CDF search \cite{Aaltonen:2011sg}, 
provided that $M_Z'$ is not so small that the jets overlap.
Masses below $M_Z/2$ are excluded by measurements of hadronic $Z$ decays. 
In the  46--77 GeV mass range there is no limit for a quark decaying into three jets; although this gap should be explored by future 
searches, we do not discuss it further.

New physics could lead to decays of vector-like fermions that are harder to detect.
If the new fermions do not mix with SM ones, and there are 
4-fermion operators  (induced, {\it e.g.,}  by a very heavy boson), 
then the vector-like fermions can decay predominantly into three SM fermions. 
These can all be light quarks if the vector-like fermion is a color triplet. The $6j$ CDF search has set a mass limit on gluinos
of 140 GeV, but the cross section for quark pair-production is smaller by a factor of about 3, so that the lower mass limit on vector-like 
quarks is only 100 GeV. Similarly, there are no limits on vector-like quarks from existing $6j$ CMS searches \cite{Chatrchyan:2013gia}.

Vector-like leptons might partially evade various collider searches if they decay into $\tau j j$. A LEP search for a pair of $\tau j$ resonances
sets a mass limit of 98 GeV on leptoquarks \cite{Abbiendi:2003iv}; the extra jet from the vector-like lepton decay would relax this limit, but it seems unlikely that 
it would be pushed below $\sim 90$ GeV.

\bigskip

{\it Quarkonium decays.}---Searches for nonstandard $\Upsilon$ decays constrain the $Z'$ couplings to $b$ quarks.  
The ratio of branching fractions $\Delta \!R_\Upsilon \equiv B( \Upsilon\! \to Z^{\prime *},\gamma^*\!  \to jj  )/B( \Upsilon\! \to \mu^+ \mu^-)$
can be used \cite{Carone:1994aa,Aranda:1998fr} to set limits on $g_z $. 
The limit on the non-electromagnetic di-jet decay of $\Upsilon (1S)$ \cite{Albrecht:1986ec}, $\Delta \!R_\Upsilon < 2.1$, gives  
the excluded region shown in Fig.~1 (labelled ``$\Upsilon$") for the $Z'_B$ model. 

The axial $Z'$ coupling to $b$ quarks 
is constrained for masses below 7 GeV by the process $\Upsilon \to \gamma Z'$. 
The search  \cite{Lees:2011wb} for $\Upsilon (2 S) \to \gamma A^0$ where $A^0$ is a pseudo-scalar decaying into hadrons, set a 
branching fraction limit of $10^{-6}$ at $M_{A^0} = 1$ GeV.
A similar limit applies to $\Upsilon \to \gamma Z'$, with the difference arising from the acceptance, which depends on the spin. 
This does not affect $Z'$s that have only a vector coupling, such as $Z'_B$. 

Charmonium decays into hadrons set limits on the $Z'$ couplings to $c$ quarks. 
We focus on exclusive $J/\psi$ and $\psi(2S)$ decays into $K^+K^-$. The photon contribution 
appears to saturate the measured 
branching fractions, although there are uncertainties from 
interference with QCD effects \cite{Czyz:2009vj}. 
The ratio $r_\psi \equiv B( \psi\! \to\!  Z^{\prime *} ,\gamma^* \! \to K^+ K^- )/B( \psi\! \to \gamma^* \! \to K^+ K^-)  $ is then bounded on both sides: 
$1/2 \sim r_\psi^{\rm min} < r_\psi \!< r_\psi^{\rm max}  \!\sim\! 2$. Computing the $Z'$ and $\gamma$ contributions in the $Z'_B$ model we obtain
\be
g_z^2  \lesssim  \frac{24 e^2}{\varepsilon_s}  \, \Big( 1-  \sqrt{r_\psi^{\rm m}}\; \Big) \left(1 - \frac{M_{Z'}^2}{M_\psi^2} \right) ~~,
\label{J-psi}
\ee
where $r_\psi^{\rm m} = r_\psi^{\rm min}$ for $M_{Z'} < M_\psi $, and $r_\psi^{\rm m} = r_\psi^{\rm max}$ for $M_{Z'} > M_\psi $;
$\varepsilon_s \approx m_s/\Lambda_{\rm QCD}$ parametrizes the flavor $SU(3)$ violation,
and $e$ is the electromagnetic gauge coupling. 
The excluded regions are shown in Fig.~1 (labelled ``$\psi$") for  $\varepsilon_s = 1/3$, with excisions at $|M_Z' /M_\psi - 1| < 10^{-2}$
where the mixing between $\psi$ and $Z'$ is large.
The constraints from inclusive decays are estimated in \cite{Aranda:1998fr}. In models with 
$z_u\neq z_d$ there are stronger constraints from $\psi\! \to\!  Z^{\prime *}  \! \to \pi^+ \pi^-$.

Similarly, the $Z'$ couplings to $s$ quarks contribute to $\phi$ meson decays into $\pi^+\pi^-$, where photon exchange dominates \cite{Oller:1999ag}.
In the $Z'_B$ model the effect violates isospin so it is suppressed by $(m_d-m_u)/\Lambda_{\rm QCD}$, and the limits are not competitive (while they are stringent
for $z_u\neq z_d$).
 
\medskip

{\it Electroweak observables.}---As long as Tr$(zY)=0$, the 1-loop kinetic mixings of $Z'$ with 
$Z$ ($c_Z$) and $\gamma$ lead only to mild constraints. The largest effect identified in 
\cite{Carone:1994aa,Aranda:1998fr} is a change in the hadronic $Z$ width.
The $Z'_B$ model with $g_z$ normalized as in Eq.~(\ref{couplings}) gives
$c_Z  \approx 0.01 g_z$  \cite{Graesser:2011vj} and 
\be
\frac{ \Delta \Gamma_Z^{\rm had}}{ \Gamma_Z^{\rm had}} = 
\frac{2 g_z c_Z c_W s_W \, (2V_u + 3 V_d)}{3 g \left( 1\!-\! M^2_{Z'}/ M^2_Z \right) \left(  2 V_u^2+ 3 V_d^2 +5/16\right)}    ~~ ,
\ee
where $V_{u,d} = \! \pm 1/4 - (3\! \pm \! 1) s^2_W/6$, and $s_W \equiv \sin {\theta_W}$.
This rules out the region labelled ``$Z$ width"  in Fig.~1. 

\bigskip

\begin{table}[b]
\renewcommand{\arraystretch}{1.2}
\begin{tabular}{c|ccc|c}\hline 
field  & $SU(3)_C$  & $SU(2)_W$   &  $U(1)_Y$  & \  $U(1)_z$ \ \ \ \\  [-0.03em] \hline
$d_R,s_R$ & 3 & \  1 \ & \  $\!\!-1/3$  &   \ $+1$ , $-1$  \ \ \ \\ [0.1em]  
$f_L,f_L'$   \ &  1 &      1     &   $\!\!+1$     &   \ $0$  \ \ \ \\
$f_R,f_R'$  \  & 1 &      1      &   $\!\!+1$    &   \ $+1$ , $-1$  \ \ \ \\ [0.1em]   
\hline
$\varphi$ & 1 & \  1 \ & \  0 \  & $+1$  \ \ \ \\ \hline
\end{tabular}
\medskip \\
\caption{\small Fields carrying $U(1)_z$ charge in the $Z'_{ds}$ model.  }
\label{table:U1d}
\end{table}

{\it Down-strange $Z'$ at the GeV scale.}---The bounds from $\Upsilon $ and $J/\psi$ decays are avoided if 
the $Z'$ couplings to $b$ and $c$ quarks vanish. This is consistent with quark mass generation, without inducing tree-level FCNC, 
when only right-handed quarks carry $U(1)_z$ charges. If $d_R$ and $s_R$ have opposite charges ($z_{d_1} = - z_{d_2}$)
and all other SM fields are $U(1)_z$ neutral (we refer to this assignment as the $Z'_{ds}$ model), then 
the only anomaly that remains to be cancelled  is  
$U(1)_Y[U(1)_z]^2$. A set of vector-like fermions that achieve that is included in Table~\ref{table:U1d}.

Tree-level  FCNC are absent provided the gauge and mass eigenstates coincide for right-handed down-type quarks.
Let us outline a mechanism for quark mass generation that satisfies this condition.
The $b$ quark acquires mass from a SM Yukawa term $ y_b H \overline b_R Q_L^3$, where 
$H$ is the Higgs doublet. 
We have chosen a basis where the Yukawa couplings of $Q_L^1$ and $Q_L^2$ to $b_R$ vanish.
The $s$ and $d$ quark masses are generated by dimension-5 operators:
\be
\frac{c^s_j}{m^\omega_1} \varphi^\dagger H \overline s_R Q_L^j  + \frac{c^d_j}{m^\omega_2} \varphi H \overline d_R Q_L^j  + {\rm H.c.}  ~~,
\label{dim-5-ops}
\ee
where $m^\omega_1$, $m^\omega_2$ are mass parameters, and $c^s_j$, $c^d_j$ are dimensionless coefficients.
Without loss of generality we take $c^s_1 =0$, so that 
$m^\omega_1$  is related to the strange quark mass by
$m^\omega_1 = c^s_2 \, v M_{Z'}/(g_z m_s) $.  For 
$c^s_2 \lesssim O(1)$ we find $m_1^\omega  \lesssim 2.5 \; {\rm TeV}/g_z$ for $M_{Z'} \approx 1 $ GeV, which
shows that the mass scale where the strange quark mass is generated may be explored at the LHC. 

A renormalizable origin of the operators (\ref{dim-5-ops}) is provided by 
two vector-like quarks, $\omega^i$, $i = 1,2$, which transform as 
$b_R$ and have Yukawa couplings to the $d$ and $s$ quarks:
\be
  y^\omega_{ij} \, H  \,  \overline \omega_R^i  \, Q_L^j   +  \lambda^s_i \, \varphi^\dagger \,\overline s_R  \, \omega_L^i
  + \lambda^d_i \,  \varphi  \,  \overline d_R  \, \omega_L^i 
  + {\rm H.c.}  
\ee
An $U(2)_L$ transformation of $Q_L^1$ and  $Q_L^2$ can set  $y^\omega_{21} = 0$.
Similarly, the $\overline b_R \omega_L^i $  terms can be rotated away, and the vector-like quark masses  
can be diagonalized, $m^\omega_i \,  \overline  \omega^i  \omega^i$. 
Comparing the operators (\ref{dim-5-ops}) with the above Yukawa  terms, we find
$c_i^s \simeq y_{1i}^\omega $ and $c_i^d \simeq  y_{2i}^\omega    \, \lambda^d_2 $ for $m^\omega_1 \ll m^\omega_2$.

\begin{figure}[t]  
\begin{center} \hspace*{-0.3cm}\includegraphics[width=0.475\textwidth, angle=0]{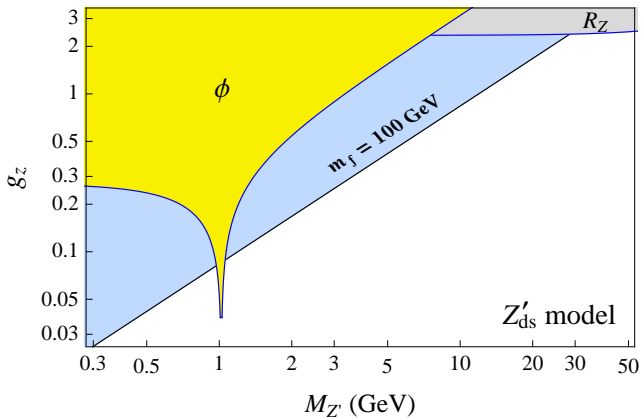}
\caption{Limits in the $g_z$ versus $M_{Z^\prime}$ plane for $Z^\prime_{ds}$.  The shaded regions are excluded (see caption to Fig 1). \\ [-8mm] 
} \label{fig:Zdslimits}
\end{center}
\end{figure}

Mass terms involving $f_L$ or $f_L'$  (see Table I) and right-handed SM leptons can be kept small by an approximate discrete symmetry.
These terms induce mixing of the vector-like fermions with the SM leptons, so that $f$ and $f'$ decay predominantly into $W \nu$. 
The LEP bound on $f$ and $f'$ masses is then around 100 GeV. 
Assuming that the form factors for vector and axial couplings are equal, 
the $g_z$ limit from $\phi$ meson decay is given by Eq.~(\ref{J-psi}) with $\varepsilon_s \to 9$ and $\psi \to\phi$, and is shown in Fig.~2. 
If $f$ and $f'$ have equal mass, then the $Z$-$Z'$ mixing is suppressed. The change in hadronic $Z$ decays,
due to $Z'$ emission and 1-loop corrections, excludes the region labelled $R_Z$ in Fig.~2 (based on \cite{Carone:1994aa} with updated  $\Delta \alpha_s$).

Additional constraints for $M_Z' \lesssim 0.5$ GeV arise from meson decays such as $\eta \! \to \gamma Z' \! \to \gamma \pi^+\pi^-$ or $\eta^\prime \! \to Z' Z' \! \to \pi^+\pi^- \pi^+\pi^- $.
Currently, the limits are weaker than the ones shown in Fig.~2, but future searches in these channels for $\pi^+\pi^-$ resonances 
may probe lower values of $g_z$.

\bigskip

{\it Conclusions.}---Besides the usual limits on light leptophobic $Z'$ bosons, from $\Upsilon$ decays and electroweak observables,
we have found a strong constraint from  the requirement that new fermions cancel the gauge anomalies.
Collider limits on the new fermion masses imply an upper limit on the gauge coupling. The same applies to other 
light gauge bosons with anomalous charges for SM fermions.

Nevertheless, the gauge coupling of a baryonic $Z'$ may be relatively large,  of order 0.1 for $M_Z' \gtrsim 2$ GeV (Fig.~1). Furthermore,
we have presented a renormalizable model where the only SM fields charged under the new group are $d_R$ and $s_R$, 
allowing even larger couplings (Fig. 2).

Future experiments may search for GeV-scale leptophobic $Z'$s in various ways, including non-standard meson decays and
LHC signatures of boosted di-jet resonances, as well as  test them through searches for
vector-like fermions.  If the  $Z'$ also couples to light dark matter 
particles, then interesting phenomena may be uncovered in neutrino detectors \cite{Batell:2009di} and other experiments. 

\bigskip

{\it Acknowledgments:} We thank Prateek Agrawal, Andre de Gouvea, Andreas Kronfeld and Felix Yu, for constructive comments.

 \vfil 

\begin{thebibliography}{99} 


\bibitem{Beringer:1900zz} 
  J.~Beringer {\it et al.}  [Particle Data Group],
  ``Review of Particle Physics",
  Phys.\ Rev.\ D {\bf 86}, 010001 (2012).

\bibitem{Dobrescu:2013cmh} 
  B.~A.~Dobrescu, F.~Yu,
  ``Coupling--mass mapping of di-jet peak searches,''
  Phys.\ Rev.\ D {\bf 88}, 035021 (2013)
  [arXiv:1306.2629].
  
\bibitem{NelsonTetradis} 
  A.~E.~Nelson, N.~Tetradis,
 ``Constraints on a new vector boson coupled to baryons,''
  Phys.\ Lett.\ B {\bf 221}, 80 (1989);

\bibitem{Krnjaic:2011ub} 
  G.~Z.~Krnjaic,
  ``Very light axigluons and the top asymmetry,''
  Phys.\ Rev.\ D {\bf 85}, 014030 (2012)
  [arXiv:1109.0648].

\bibitem{Fox:2011qd} 
  P.~J.~Fox  {\it et al.},  
  ``An effective $Z'$,''
  Phys.\ Rev.\ D {\bf 84}, 115006 (2011)
  [arXiv:1104.4127].
  
\bibitem{Carone:1994aa} 
  C.~D.~Carone, H.~Murayama,
  ``Possible light $U(1)$ gauge boson coupled to baryon number,''
  Phys.\ Rev.\ Lett.\  {\bf 74}, 3122 (1995)
  [hep-ph/9411256];
  ``Realistic models with a light $U(1)$ gauge boson coupled to baryon number,''
  Phys.\ Rev.\ D {\bf 52}, 484 (1995)
  [hep-ph/9501220];
  D.~C.~Bailey, S.~Davidson,
  ``Is there a vector boson coupling to baryon number?'',
  Phys.\ Lett.\ B {\bf 348}, 185 (1995)
  [hep-ph/9411355].

\bibitem{Aranda:1998fr} 
  A.~Aranda, C.~D.~Carone,
  ``Limits on a light leptophobic gauge boson,''
  Phys.\ Lett.\ B {\bf 443}, 352 (1998)
  [hep-ph/9809522].

\bibitem{Bardeen:1969md} 
  W.~A.~Bardeen,
  ``Anomalous Ward identities in spinor field theories,''
  Phys.\ Rev.\  {\bf 184}, 1848 (1969).

\bibitem{Carena:2004xs} 
  M.~S.~Carena, {\it et al.},  
  ``$Z^\prime$ gauge bosons at the Tevatron,''
  Phys.\ Rev.\ D {\bf 70}, 093009 (2004)
  [hep-ph/0408098].

\bibitem{Chatrchyan:2013uxa} 
  S.~Chatrchyan {\it et al.}  [CMS Collaboration],
  ``Inclusive search for a vector-like $T$ quark with charge $2/3$", 
  Phys.\ Lett.\ B {\bf 729}, 149 (2014)
  [arXiv:1311.7667];
  ``Combined search for the quarks of a sequential fourth generation,''
  Phys.\ Rev.\ D {\bf 86}, 112003 (2012)
  [arXiv:1209.1062].

\bibitem{deGouvea:2012hc} 
  A.~de Gouvea, W.~-C.~Huang, J.~Kile,
  ``Dark matter from weak polyplets,''
  arXiv:1207.0510.

\bibitem{LongLived}
ATLAS Collaboration,
``A search for heavy long-lived sleptons using 16 fb$^{-1}$ of $pp$ collisions",
 note CONF-2013-058, Aug. 2013.

\bibitem{Alwall:2011uj} 
  J.~Alwall, {\it et al.}, 
  ``MadGraph 5: Going Beyond,''
  JHEP {\bf 1106}, 128 (2011)
  [arXiv:1106.0522].

\bibitem{Thomas:1998wy} 
  S.~D.~Thomas, J.~D.~Wells,
  ``Phenomenology of massive vectorlike doublet leptons,''
  Phys.\ Rev.\ Lett.\  {\bf 81}, 34 (1998)
  [hep-ph/9804359];
  S.~Dimopoulos, {\it et al.}   
  ``TeV dark matter'',
  Nucl.\ Phys.\ B {\bf 349}, 714 (1991)
  [Erratum-ibid.\ B {\bf 357}, 308 (1991)].

\bibitem{Heister:2002mn} 
  A.~Heister {\it et al.}  [ALEPH Collaboration],
  ``Search for charginos nearly mass degenerate with the lightest neutralino", 
  Phys.\ Lett.\ B {\bf 533}, 223 (2002)
  [hep-ex/0203020].

\bibitem{Halverson:2014nwa} 
  J.~Halverson, N.~Orlofsky, A.~Pierce,
  ``Vectorlike leptons as the tip of the dark matter iceberg,''
  arXiv:1403.1592.
 
\bibitem{Duerr:2013dza} 
  M.~Duerr, P.~Fileviez Perez, M.~B.~Wise,
  ``Gauge theory for baryon and lepton numbers with leptoquarks,''
  Phys.\ Rev.\ Lett.\  {\bf 110},  231801 (2013)
  [arXiv:1304.0576].
 
\bibitem{Falkowski:2013jya} 
  A. Falkowski, D.M. Straub, A. Vicente,
  ``Vector-like leptons: Higgs decays and collider phenomenology'',
  arXiv:1312.5329.
 
\bibitem{Achard:2001qw} 
  P.~Achard {\it et al.}  [L3 Collaboration],
  ``Search for heavy neutral and charged leptons in $e^{+} e^{-}$ annihilation at LEP,''
  Phys.\ Lett.\ B {\bf 517}, 75 (2001)
  [hep-ex/0107015].
    
\bibitem{Aaltonen:2011sg} 
  T.~Aaltonen {\it et al.}  [CDF Collaboration],
  ``First search for multijet resonances in $\sqrt{s} = 1.96$ TeV $ p\bar{p}$ collisions,''
  Phys.\ Rev.\ Lett.\  {\bf 107}, 042001 (2011)
  [arXiv:1105.2815].
  
\bibitem{Chatrchyan:2013gia} 
  S.~Chatrchyan {\it et al.}  [CMS Collaboration],
  ``Searches for light- and heavy-flavour three-jet resonances  at $\sqrt{s} = 8$ TeV,''
  Phys.\ Lett.\ B {\bf 730}, 193 (2014)
  [arXiv:1311.1799];
  ``Search for three-jet resonances  at $\sqrt{s}=7$ TeV,''
  Phys.\ Lett.\ B {\bf 718}, 329 (2012)
  [arXiv:1208.2931].
  
\bibitem{Abbiendi:2003iv} 
  G.~Abbiendi {\it et al.}  [OPAL Collaboration],
  ``Search for pair produced leptoquarks in $e^{+} e^{-}$ interactions", 
  Eur.\ Phys.\ J.\ C {\bf 31}, 281 (2003)
  [hep-ex/0305053].
      
\bibitem{Albrecht:1986ec} 
  H.~Albrecht {\it et al.}  [ARGUS Collaboration],
  ``An upper limit for two jet production in direct $\Upsilon (1s)$ decays,''
  Z.\ Phys.\ C {\bf 31}, 181 (1986).
 
\bibitem{Lees:2011wb} 
  J.~P.~Lees {\it et al.}  [BaBar Collaboration],
  ``Search for hadronic decays of a light Higgs boson in the radiative decay $\Upsilon \to \gamma A^0$,''
  Phys.\ Rev.\ Lett.\  {\bf 107}, 221803 (2011)
  [arXiv:1108.3549].

\bibitem{Czyz:2009vj} 
  H.~Czyz and J.~H.~Kuhn,
``Strong and electromagnetic $J/\psi$ and $\psi (2S)$ decays into pion and kaon pairs,''
  Phys.\ Rev.\ D {\bf 80}, 034035 (2009); [arXiv:0904.0515].

\bibitem{Oller:1999ag} 
  J.~A.~Oller, E.~Oset, J.~R.~Pelaez,
  ``The $\phi \to \pi^+ \pi^-$ decay within a chiral unitary approach,''
  Phys.\ Rev.\ D {\bf 62}, 114017 (2000)
  [hep-ph/9911297].

\bibitem{Graesser:2011vj} 
  M.~L.~Graesser, I.~M.~Shoemaker, L.~Vecchi,
  ``A dark force for baryons,''
  arXiv:1107.2666.

\bibitem{Batell:2009di} 
  B.~Batell, M.~Pospelov, A.~Ritz,
  ``Exploring portals to a hidden sector through fixed targets,''
  Phys.\ Rev.\ D {\bf 80}, 095024 (2009)
  [arXiv:0906.5614].

\end{thebibliography}
\end{document}